\documentclass[aps,twocolumn,showpacs,showkeys,preprintnumbers,amsmath,amssymb]{revtex4}
\usepackage{epsfig}

\usepackage{dcolumn}
\usepackage{amsmath}

\usepackage{graphicx}
\usepackage{rotating}
\usepackage{multirow}

\def\be{\begin{equation}}
\def\ee{\end{equation}}
\def\lsim{\raise0.3ex\hbox{$<$\kern-0.75em\raise-1.1ex\hbox{$\sim$}}}
\def\gsim{\raise0.3ex\hbox{$>$\kern-0.75em\raise-1.1ex\hbox{$\sim$}}}

\begin{document}

\newlength{\figurewidth}
\ifdim\columnwidth<10.5cm
  \setlength{\figurewidth}{0.95\columnwidth}
\else
  \setlength{\figurewidth}{10cm}
\fi
\setlength{\parskip}{0pt}
\setlength{\tabcolsep}{6pt}
\setlength{\arraycolsep}{2pt}

\title{Number of spanning clusters at the high-dimensional percolation
thresholds}
\medskip

\author{Santo Fortunato}

\vskip0.3cm

\affiliation{Fakult\"at f\"ur Physik, Universit\"at Bielefeld,
D-33501 Bielefeld, Germany}

\author{Amnon Aharony}

\vskip0.3cm

\affiliation{School of Physics and Astronomy, Raymond and Beverly Sackler
Faculty of Exact Sciences, Tel Aviv University, Tel Aviv 69978,
}

\affiliation{Department of Physics, Ben Gurion University, Beer Sheva 84105, Israel}

\author{Antonio Coniglio}

\vskip0.3cm

\affiliation{Dipartimento di Scienze Fisiche, Universit\`a di Napoli `Federico II'
and Unit\'a INFM-Coherentia, Via Cintia, I-80126 Naples, Italy}

\author{Dietrich Stauffer}

\vskip0.3cm

\affiliation{Institute for Theoretical Physics, Cologne University, D-50923
  K\"oln, Germany}

\begin{abstract}
\noindent A scaling theory is used to derive the dependence of the
average number ${\langle}k{\rangle}$ of spanning clusters at
threshold on the lattice size $L$. This number  should become
independent of $L$ for dimensions $d<6$, and vary as log $L$ at
$d=6$.
The predictions for $d>6$ depend on the boundary
conditions, and the results there may vary between $L^{d-6}$ and
$L^0$.
While simulations in six dimensions are consistent with
this prediction (after including corrections of order $\log(\log
L)$), in five dimensions the average number of spanning clusters
still increases as log $L$ even up to $L = 201$. However, the
histogram $P(k)$ of the spanning cluster multiplicity does scale
as a function of $k X(L)$, with $X(L)=1+{\rm const}/L$, indicating
that for sufficiently large $L$ the average $\langle k \rangle$
will approach a finite value: a fit of the 5D multiplicity data
with a constant plus a simple linear correction to scaling
reproduces the data very well. Numerical simulations for $d>6$ and
for $d=4$ are also presented.
\end{abstract}



\maketitle

\vskip0.7cm

\section{Introduction and Theory}

We are interested in site percolation on a finite hypercubic
lattice in $d$ dimensions, of linear size $L$. The number of
spanning clusters when $L$ is of the order of the percolation
correlation length, $\xi$, away from the percolation threshold
$p_c$, has been discussed originally in \cite{ac}. The purpose of
this paper is to discuss the average number of spanning clusters,
$\langle k \rangle$, {\it at}  $p=p_c$, when $\xi$ is infinite,
and all the critical quantities behave as powers of $L$.
As we discuss below, the number of spanning clusters has been the
topic of much discussion in the literature. For example, although
intuitively one would think that in two dimensions there exists
only one spanning cluster, in fact (for appropriate boundary
conditions) there is a whole distribution of the sizes of such
clusters \cite{aizenman}. In addition, we show below that for
$d<6$ the average number of such clusters is finite, in direct
relation with the validity of the hyper-scaling relations among
critical exponents. For $d>6$ hyper-scaling is violated, dangerous
irrelevant variables must be introduced, and the result for
$\langle k \rangle$ becomes ambiguous, depending on the boundary
conditions. In fact, the theory for that case is yet incomplete,
leaving an open challenge for future research.

 We start with a theoretical discussion.
The percolation ``order parameter" $P_{span}$ is usually defined
as the probability that a site belongs to {\it any} spanning
cluster (i.e. to the union of {\it all} spanning clusters). For a
finite sample, $P_{span}$ is related to the cluster size
distribution function $n_s(p,L)$ (defined as the average number
per site of clusters containing $s$ sites) via the sum rule
\cite{book}
\begin{equation}
P_{span}(p,L)=p-\sum_s s n_s(p,L), \label{pspan}
\end{equation}
where the sum over $s$ excludes all clusters which span the lattice.
The exact details depend on the definition of ``spanning", e.g.
along how many directions should the cluster connect opposite
faces of the hypercube. However, these details do not matter for
the scaling arguments presented below.

For a finite lattice, the sum in Eq. (\ref{pspan}) goes up to
$s_{max}(L)$, which is of the same order as the average mass of a
single spanning cluster, $s(L)$. Since the total mass of {\it all}
spanning clusters is given by $L^dP_{span}$, this implies that the
average number ${\langle}k{\rangle}$ of spanning clusters is given by
\begin{equation}
{\langle}k{\rangle} \propto L^d P_{span}/s(L). \label{k}
\end{equation}
This relation should hold at all dimensions.
The proportionality constant in Eq. (\ref{k}) (which results,
among other things, from dividing averages rather than averaging
the ratio and on the detailed definition of ``spanning" in a
finite sample) may depend on the boundary conditions and on other
details.

One way to derive $s(L)$ is to use the pair connectivity function,
$G(r)$. This function yields the probability that a site at
distance $r$ from the origin is connected to the same cluster as
the origin. Assuming that the site at the origin belongs to any
spanning cluster, with probability $P_{span}$, the density profile
of the cluster is given by $\rho(r)=G(r)/P_{span}$
\cite{essam,kapitulnik}. Thus,
\begin{equation}
s(L) \propto \int_0^{L} d^dr \; G(r)/P_{span}=S(L)/P_{span},
\end{equation}
where $S(L)=\int_0^{L} d^dr \; G(r)$ is
proportional
 to the mean cluster size,
$$S \propto \sum_s s^2n_s(p,L)/\sum_s sn_s.$$
(These two quantities are equal in the thermodynamic limit, $L
\rightarrow \infty$.)
Thus, we come to the fundamental relation
\begin{equation}
{\langle}k{\rangle} \propto L^d P_{span}^2/S(L). \label{kk}
\end{equation}

A naive finite size scaling theory would predict that at $p=p_c$
one has $P_{span} \propto L^{-\beta/\nu}$ and $S(L) \propto
L^{\gamma/\nu}$ \cite{book}. Thus, one concludes that
\begin{equation}
{\langle}k{\rangle} \propto L^{d-(2\beta+\gamma)/\nu}.
\end{equation}
For $d<6$, hyper-scaling implies that $d\nu=2\beta+\gamma$, and
therefore ${\langle}k{\rangle}$ is asymptotically independent of
$L$. In fact, the combination of amplitudes which appears in Eq.
(\ref{k}) is universal, depending only on the type of boundary
conditions \cite{privman}. As we discuss below, numerical
estimates for ${\langle}k{\rangle}$ indeed approach a constant for
$d \le 4$. However, data for $d=5$ require further discussion.
Although not a major purpose of this paper, our discussion below
should also serve as a warning and as a guideline for future
numerical simulations in such high dimensions. In fact, some data
remain ambiguous unless one uses available theoretical
information, or unless one performs simulations on much larger
scales than presently possible.

\begin{figure}[htb]
\begin{center}
\resizebox{\figurewidth}{!}{\includegraphics[angle=0]{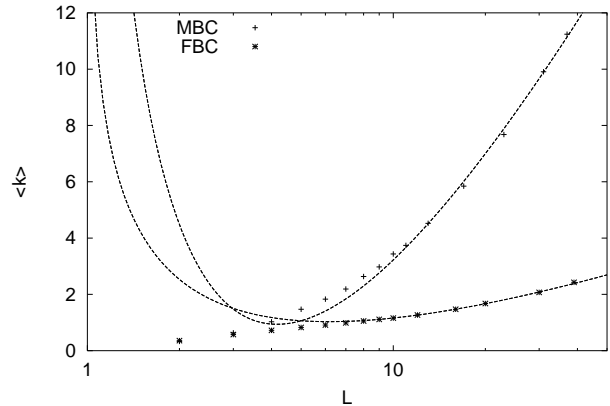}}
\end{center}
\caption{Average number ${\langle}k{\rangle}$ of simultaneously
spanning clusters at the six-dimensional percolation threshold,
for FBC (*) and MBC (+) (see text). The lines are fits to  ${\langle}k{\rangle}= A
\ln(L) +B \ln(\ln L)$; we used the data of \cite{coniglio} plus
additional simulations for small $L$ with the number of samples
variable from 1000 to 50000.}
\label{fig1}
\end{figure}

At $d=6$, many power laws are modified by logarithmic corrections.
In fact \cite{aharony},
\begin{eqnarray}
P_{span} \propto L^{-2}(\ln L)^{11/21},\nonumber\\
S(L) \propto L^2(\ln L)^{1/21}.
\end{eqnarray}
Using our basic result, Eq. (\ref{kk}),
 we thus find
\begin{equation}
{\langle}k{\rangle} \propto \frac{L^6 P_{span}^2}{S(L)} \propto \ln L,\ \ \ \ d=6.
\end{equation}
Further analysis shows that the coefficient of proportionality
here (and also the result for ${\langle}k{\rangle}$ for $d<6$) is a universal number
\cite{privman}.  Corrections to the above result will involve
$\ln(\ln L+$ {\rm const}). Fig. \ref{fig1} shows that the data
in six dimensions, which were fitted by the square of log $L$ in
\cite{coniglio}, can also be fitted to a simple logarithm plus a
log log $L$ correction to scaling, ${\langle}k{\rangle}=A \ln L+B \ln \ln L+\dots$.
The two data sets refer respectively to free boundary conditions
(FBC) and to mixed boundaries (MBC), i.e. helical in $d-2$
directions and free in the remaining two. In both cases we obtain
a good fit for $L>10$, so that the trend appears independent of
boundary conditions, as one would expect. The coefficients of the
fits are  $A=11.9,~B=-17$ (MBC) and $A=2.38,~B=-4.33$  (FBC).

For $d>6$, hyper-scaling is broken, and one has the mean-field
exponents $\beta=\gamma=2\nu=1$. A simple scaling in which $\xi$
is replaced by $L$ \cite{ac} would give $P_{span} \propto
L^{-2},~S(L) \propto L^2,~ s(L) \propto L^4$ and
\begin{equation}
{\langle}k{\rangle} \propto L^{d-6}. \label{kkk}
\end{equation}
This simple result, which already appeared in \cite{book}, has
also been proved by Aizenman \cite{aizenman} for the case of bulk
boundary conditions, where a spanning cluster connects two
opposite faces of the box of size $L$ under the condition that
sites in the box can also be connected by paths outside the box.

 However, other forms of scaling may be possible,
which may depend on the boundary conditions. We will discuss the
different scaling approaches within the framework of the
renormalization group (RG) theory (e.g. Ref. \cite{aharony}). This
theory is conveniently discussed using the ``free energy" $F$,
equal to the generating function of the cluster distribution
function,
\begin{equation}
F(p,h,L)=\sum_{s}n_s(p,L)e^{-sh},
\end{equation}
which is related to the free energy of the $q$-state Potts model
for $q \rightarrow 1$. $P_{span}$ and $S(L)$ are then the first
and second derivatives of the singular part of $F$ with respect to
$h$. After $\ell$ RG iterations, this singular part becomes
\begin{equation}
f(p,h,w,L)=e^{-d\ell}f(t(\ell),h(\ell),w(\ell),L/e^\ell),
\end{equation}
where $t=p-p_c$. Here, $w$ represents the probability for having a
three-fold vertex at a site on a cluster.
Although irrelevant in the RG sense, this variable must be
included in the analysis, since the ``free energy" may depend on
it in a singular way, causing the breakdown of hyper-scaling. This
is why $w$ is called ``a dangerous irrelevant variable".
For a
finite sample close to $p_c$, $L \ll \xi$, iteration until
$e^\ell=L$ yields the $L$-dependence of the quantities discussed
above.

For $d<6$, $w(\ell)$ approaches a finite fixed point value, and
one ends up with hyper-scaling and with the conclusion that $k$
approaches a finite constant. For $d>6$, one ends up with
\begin{equation}
f(t,h,w,L)=L^{-d}f(tL^2,hL^{d/2+1},wL^{3-d/2},1). \label{f1}
\end{equation}
However, as stated above, now $w$ turns out to be a ``dangerous
irrelevant variable". One way to see this is to consider the
infinite sample. After eliminating the fluctuations, $f$ is given
by the minimum of the Landau free energy,
\begin{equation}
f=\min_P[t P^2+w P^3-h P],
\end{equation}
while the order parameter $P_{span}$ is equal to the value of $P$
which minimizes this free energy. Replacing $P$ by $Qw^{-1+x/3}$,
with an arbitrary exponent $x$, it is easy to see that $f$ obeys
the scaling relation
\begin{equation}
f(t,h,w)=w^{-2+x}f(tw^{-x/3},hw^{1-2x/3},1). \label{f3}
\end{equation}
In particular, the minimization with respect to $Q$ now yields the
equation $3Q^2+2Qtw^{-x/3}-hw^{1-2x/3}=0$, leading to the scaling
form
\begin{equation}
f(t,h,w)=w^{-2}t^3g(hw/t^2), \label{f2}
\end{equation}
which is completely independent of the arbitrary exponent $x$.
This ambiguity stems from the fact that the Landau free energy is
calculated for an infinite system.

Unlike the above result for the infinite system, the exponent $x$
does persist for {\it finite} samples. In fact, the theoretical
predictions for $\langle k \rangle$ depend crucially on $x$, and
this remains an open challenge for future research. In this case,
combining the scaling with $L$ from Eq. (\ref{f1}) with Eq.
(\ref{f3}), we end up with
\begin{eqnarray}
f(t,h,w,L)=w^{x-2}L^{-6+3x-dx/2}\times\nonumber\\
\times f(tw^{-x/3}L^{2-x+dx/6},hw^{1-2x/3}L^{4-2x+dx/3},1,1),
\label{f4}
\end{eqnarray}
and the value of $x$ must follow from the boundary condition,
which breaks the scale invariance reflected in Eq. (\ref{f2}).
Taking derivatives with respect to $h$, and setting $t=0$, this
implies that
\begin{eqnarray}
P_{span} \propto L^{-2+x-dx/6},\nonumber\\
 S \propto L^{2-x+dx/6},\nonumber\\
 s(L) \propto L^{4-2x+dx/3}.
 \end{eqnarray}
 The naive scaling result of Eq. ({\ref{kkk}) is obtained with the
 simple choice $x=0$.

\begin{figure}[htb]
\begin{center}
\resizebox{\figurewidth}{!}{\includegraphics[angle=0]{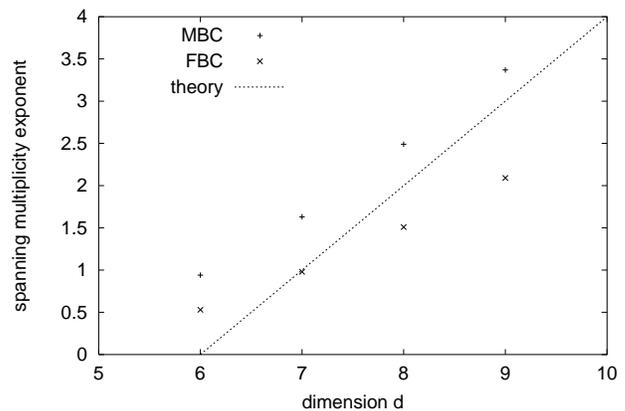}}
\end{center}
\caption{Effective exponents
for the scaling with $L$ of the spanning cluster multiplicity (from
\cite{coniglio}), as a function of the spatial dimension $d$.
The dashed line represents the theoretical prediction from Eq. (\ref{kkk}).}
\label{fig2}
\end{figure}

Fig. \ref{fig2} compares Eq. (\ref{kkk}) with the numerically
estimated effective exponents. Clearly, there exist some
discrepancies between theory and simulations. However, these
discrepancies  do not worry us much since the lattices for $d>6$
are very small in their linear dimension $L$. This is confirmed by
the fact that, for $d>6$, the results for the exponents strongly depend
on the type of boundary conditions one chooses. Also, our
attempts to add corrections to scaling to the leading term of the
fit {\it ansatz} did not improve the situation.
At present, the numerical results cannot clearly confirm that
$x=0$, as would be required if Aizenman's \cite{aizenman} result
also applies for our boundary conditions (note that his proof works
under somewhat different conditions). Thus, the value of $x$
remains to be determined in the future.

A second possibly relevant case is $x=2$. In this case, $f$ has
the form
\begin{equation}
f(t,h,w,L)=L^{-d}{\cal F}(w^{-2/3}tL^{d/3},hw^{-1/3}
L^{2d/3},1,1), \label{fff}
\end{equation}
yielding $D=2d/3$ and  $P_{span}$ and $S$ given by
\begin{equation}
P_{span} = L^{-d/3}f_1(w^{-2/3}tL^{d/3}), \label{pp}
\end{equation}
\begin{equation}
S = L^{d/3}f_2(w^{-2/3}tL^{d/3}). \label{ss}
\end{equation}
In the limits $ t\rightarrow 0$, $L \rightarrow \infty$, these
equations  behave as
\begin{equation}
P_{span}\propto S^{-1} \propto t \propto \xi^{-2},\ \ \  \quad L
=\infty, \label{pp0}
\end{equation}
\begin{equation}
P_{span} \propto S^{-1} \propto L^{-d/3},\ \ \  \quad {\langle}k{\rangle}\propto
{\rm const},\ \ \  \quad t=0, \label{pp1}
\end{equation}
and
\begin{equation}
P_{span} \propto S^{-1} \propto L^{-y},\ \ \  \quad
{\langle}k{\rangle}\propto L^{d-3y},\nonumber
\end{equation}
\begin{equation}
tL^y={\rm const} <0 ,\ \ \  \quad y\le d/3. \label{pp2}
\end{equation}
(Remember: these equations only apply for $d>6$).
The scaling behavior of Eqs. (\ref{pp0})-(\ref{pp2}) for $y=2$ is
analogous to that found by Chen and Dohm \cite{chen} for the $\phi
^4$ Ising model with periodic boundary conditions, and has
recently been used to find an upper bound for the number of
spanning clusters \cite{coniglio}. As seen from Fig. \ref{fig2},
our data for $t=0$ do not seem to agree with the prediction that
${\langle}k{\rangle}$ approaches a finite constant. However, our
samples are small, and the situation for $d>6$ requires more
studies. Very recent simulations show that $P_{span}$ differs
appreciably from the relative size of the largest cluster
\cite{zekri}, but that is not enough to conclude that the spanning
cluster multiplicity diverges for $d>6$.
Thus, the true behavior of $\langle k \rangle$ for $d>6$ remains
an open issue. We hope that the present paper will stimulate both
numerical and theoretical discussions of this question.

\section{Results of the Simulations}

We now review in detail our numerical simulations. The first numerical
studies on this topic were performed by De Arcangelis \cite{lucilla}.

\begin{figure}[htb]
\begin{center}
\resizebox{\figurewidth}{!}{\includegraphics[angle=0]{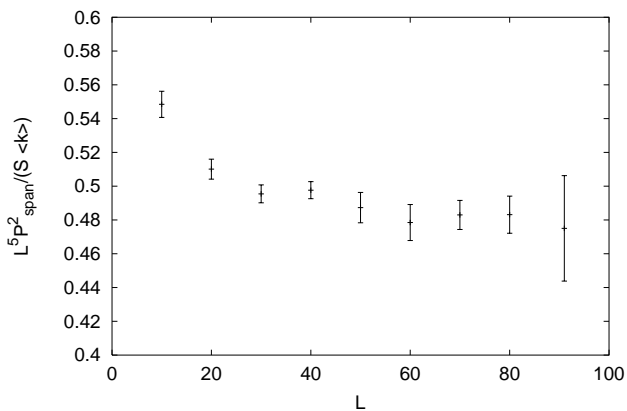}}
\end{center}
\caption{Numerical verification of our relation Eq. (\ref{kk}),
  for 5D percolation with MBC.}
\label{fig2a}
\end{figure}

An important issue concerns the validity of our Eq. (\ref{kk}), which is
a general result that establishes a link between the order parameter
$P_{span}$, the mean cluster size $S$ and the spanning cluster multiplicity ${\langle}k{\rangle}$.
Figs. \ref{fig2a} and \ref{fig2b} show numerical tests of this relation.
We analyzed one case below the upper critical dimension $d_c=6$, i. e.
5D with MBC, and one case above $d_c$, i. e. 7D with FBC.
In both cases we have calculated the ratio $L^d{P^2_{span}}/(S\,{\langle}k{\rangle})$.
From our Eq.(\ref{kk}) we expect that for $L$ large this ratio
converges to a constant, and both our figures confirm this expectation.
We stress that the $S$ that we calculate through our simulations differs
from the standard definition of $S=\sum_s s^2n_s(p,L)/\sum_s sn_s$
by the absence of the denominator. The latter is smaller than $1$,
as it is the density of occupied sites which belong to finite clusters,
so the real value of the ratio $L^d{P^2_{span}}/(S\,{\langle}k{\rangle})$ would be smaller than
the one we show in Figs. \ref{fig2a} and \ref{fig2b}.
On the other hand
the functional dependence on $L$ of the ratio is the same in both cases.

\begin{figure}[htb]
\begin{center}
\resizebox{\figurewidth}{!}{\includegraphics[angle=0]{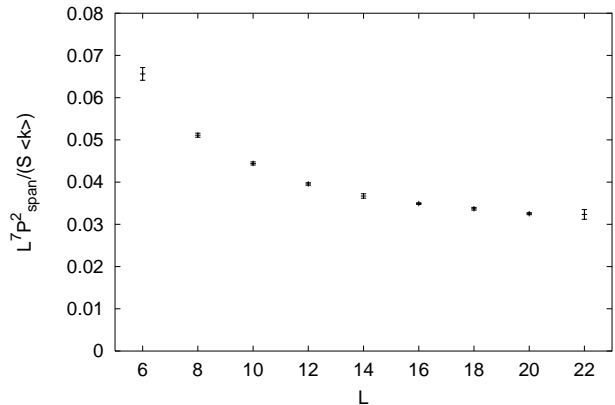}}
\end{center}
\caption{As Fig. \ref{fig2a}, but for 7D percolation with FBC.}
\label{fig2b}
\end{figure}

Let us now concentrate on the trend of the multiplicity when
the linear dimension $L$ of the lattice varies.
In less than
five dimensions, Ref. \cite{coniglio} found an asymptotically
constant number of spanning clusters for both theory and
simulation. Thus the real problem is five dimensions, which we
discuss now.

Fig. \ref{fig3} shows new data for $L^5$ sites, extending up to $L
= 201$, the largest five-dimensional system known to us from
direct simulations ($176^5$ was simulated in Ising models
\cite{stauffer}). They are still compatible with the number of
spanning clusters increasing as log $L$. These data use MBC, but
similar proportionalities to log $L$ were found also with free and
periodic boundary conditions (PBC, which means helical boundaries
in $d-1$ directions), which are illustrated in Fig. \ref{other5D}.
By looking at the data corresponding to periodic boundaries,
however, it seems that the curve smoothly bends towards the end,
which might indicate the beginning of a crossover.

\begin{figure}[htb]
\begin{center}
\resizebox{\figurewidth}{!}{\includegraphics[angle=0]{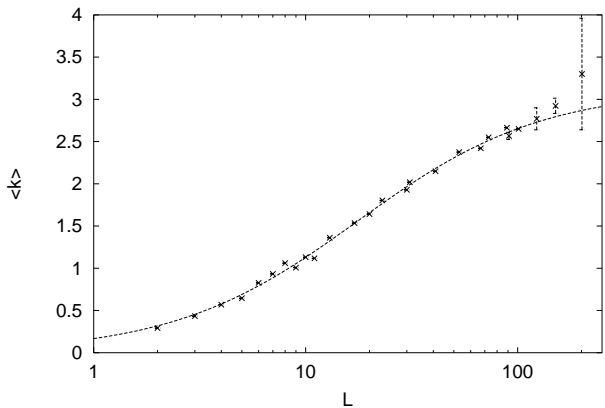}}
\end{center}
\caption{Five-dimensional problem, MBC. The
right-most error bar for $L = 201$  comes from 10 samples, the two
smaller error bars shown from 100 samples. Most of the data use
1000 samples with error bars too small to be shown. There is no
evidence for a plateau; nevertheless, the trend of the data is
beautifully reproduced just by adding a simple non-logarithmic
scaling correction (dashed line in the plot).}
\label{fig3}
\end{figure}

For comparison, Fig. \ref{4Dsol} shows four-dimensional
simulations with MBC. There we see for small $L \le 12$ a
logarithmic increase, followed by a crossover region extending
over one decade in $L$, and ending finally in the theoretically
predicted plateau near $L = 200$. Further simulations with PBC
(not shown) confirm this trend. These four-dimensional results are
a nice example for the need of large lattices in simulations: only
above $L = 100$ the theory is confirmed. Comparing Figs.
\ref{fig3} and \ref{4Dsol} we may hope that also in five
dimensions a crossover would be seen towards a plateau, if we
could simulate larger lattices than the present world records
\cite{tigg}.

\begin{figure}[htb]
\begin{center}
\resizebox{\figurewidth}{!}{\includegraphics[angle=0]{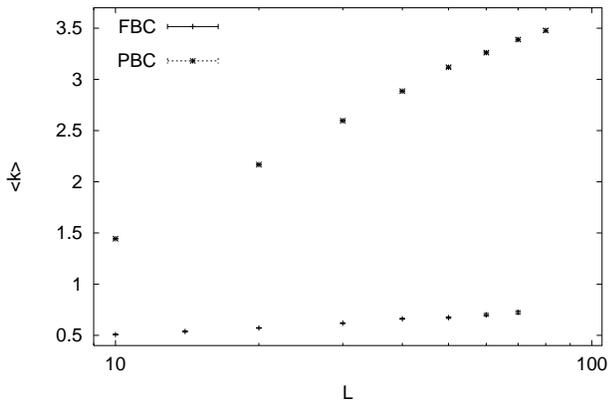}}
\end{center}
\caption{Five-dimensional data of the spanning cluster multiplicity at threshold for
FBC (+) and PBC (*). One still finds an
increase with $L$, although for PBC
it seems that the data curve starts bending at high $L$.
The number of samples goes
from 3000 to 50000 for FBC and from 10000 to 50000 for PBC.}
\label{other5D}
\end{figure}

There is a way, however, to clarify the issue.
Instead of looking merely at the
average number $\langle k \rangle$ of spanning clusters (as in all of the
above discussion), we can analyze the histogram $P(k)$ of that number. Its
tail for large $k$ was already shown to be consistent with theory \cite{shchur}:
$-\ln\,P(k \rightarrow \infty) \propto k^{d/(d-1)}$ for $2 \le d \le 7$. If
$\langle k \rangle$ would be size-independent, on a lattice
of linear dimension $L$ we would also expect $P(k,L)=P(k)$, i.e.
the histogram is as well size-independent.
Finite-size scaling for large $L$
would nevertheless allow correction factors like $X(L)=1 + {\rm const}/L$, so
that
the histogram is indeed a function $P(kX(L))$.
On the other hand, $P(kX(L))$ is normalized to some $L$-independent number $C$
(we chose here $C=1000$), and therefore
\begin{equation}
\sum_{k=0}^{\infty}P(kX(L))=C. \label{normP}
\end{equation}
If we approximate the sum with the integral over $k$, we can
perform the change of variable $j=kX(L)$, so to get
\begin{equation}
\frac{1}{X(L)}\sum_{j=0}^{\infty}P(j)=C.
\label{normPz}
\end{equation}
Equation (\ref{normPz}) cannot be right as it stands, because the
left hand side is a function of $L$, while the right hand side is
not. That means that the histogram $P(k,L)$ is not a scaling
function of the variable $j=kX(L)$, but that
$P(k,L)=X(L)P^{\prime}(kX(L))$, where $P^{\prime}$ is
now \cite{nota} a scaling function of $j$. Let
us check what happens to the average multiplicity $\langle k
\rangle$ if we assume that $P(k,L)$ has the above-derived form:
\begin{equation}
\langle k \rangle=\frac{\sum_{k=0}^{\infty}kX(L)P^{\prime}(kX(L))}
{\sum_{k=0}^{\infty}X(L)P^{\prime}(kX(L))}=\frac{1}{CX(L)}\sum_{j=0}^{\infty}jP^{\prime}(j),
\label{avmult}
\end{equation}
where we again made the substitution $j=kX(L)$ in the sum over $k$.
The sum in (\ref{avmult}) is independent of $L$, and we finally obtain:
\begin{equation}
\langle k \rangle \,{\propto}\, \frac{1}{X(L)}, \label{avmultfin}
\end{equation}
so that the whole $L$ dependence
of the average multiplicity is contained in the correction factor $X(L)$.

\begin{figure}[htb]
\begin{center}
\resizebox{\figurewidth}{!}{\includegraphics[angle=0]{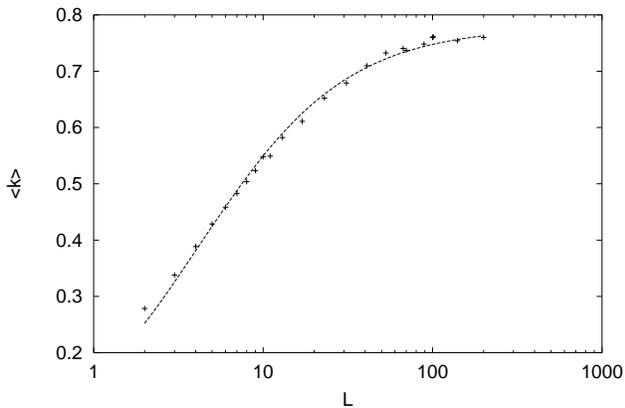}}
\end{center}
\caption{Four-dimensional solution, MBC. The spanning
cluster multiplicity increases in a wide range of $L$, but attains
finally a plateau for $L>100$. The dashed curve is the fit with
the simple correction to scaling {\it ansatz} $aL/(L+b)$. For each $L$ we took
mostly $1000$ samples.}
\label{4Dsol}
\end{figure}

In this way, we have now the chance to make a cross check on our data.
If $\langle k \rangle$ is indeed size-independent for $d<6$, we should
be able to find a simple non-logarithmic correction factor $X(L)=1+{\rm
const}/L$,
such that $\langle k \rangle\, \propto\, 1/X(L)$ and consistently
$P(k,L)/X(L)$ is a scaling function of the variable $j=kX(L)$.
We remark that only eventual discrepancies of the rescaled histograms at large
$k$
can lead to infinitely many spanning clusters in the limit
$L\rightarrow\,\infty$: if, for all $L$, there were a finite $k^*$ beyond which
the histograms collapse, $\langle k \rangle$ would necessarily approach a
constant for large
$L$, modulo finite size corrections.

We tried then to fit the multiplicity curves in four and five
dimensions with the simple two-parameters {\it ansatz} $aL/(L+b)$,
to see whether we could reproduce the observed trends. As a first
trial we took the four-dimensional data of Fig. \ref{4Dsol} and
fitted several portions of the increasing part of the curve,
before the plateau. All our fits are quite good; besides, starting
from the range $[2:60]$, the fit is stable, i.e. we obtain the
same values for the parameters $a$ and $b$ as for the fit on the
full curve, within errors. This is quite interesting, because it
allows us to predict quite precisely where saturation takes place,
even if one analyzes values of $L$ which lie well below the
beginning of the plateau. The best fit curve, for which $a=0.78$
and $b=4.2$, is plotted in Fig. \ref{4Dsol}.

\begin{figure}[htb]
\begin{center}
\resizebox{\figurewidth}{!}{\includegraphics[angle=0]{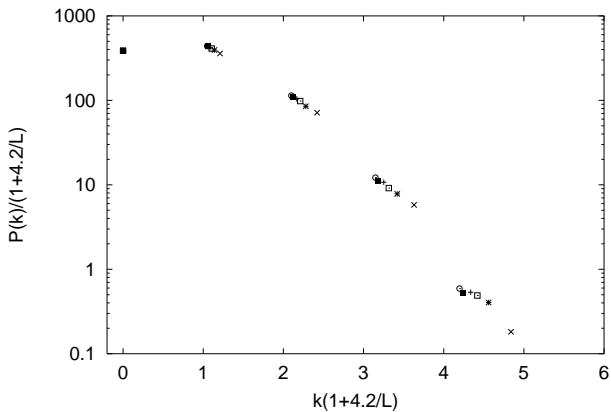}}
\end{center}
\caption{Four-dimensional histogram, MBC. Occurrence $P(k,L)$ of samples with $k$
spanning clusters
each, versus $k$, with the finite-size correction factor $1+4.2/L$
derived from the fit of the average multiplicity.
The data sets have a variable number of samples, mostly $50000$, we normalized
them all to 1000 iterations. The nice scaling is consistent with
the size-independence of the average multiplicity. The lattices we have taken
are: $20^4$, $30^4$, $40^4$, $50^4$, $70^4$, $85^4$.}
\label{histo4D}
\end{figure}

As a nice confirmation of this result, we
show in Fig. \ref{histo4D}
the corresponding histograms $P(k,L)$, where we
use the value $b=4.2$ derived above for the scaling correction.
We chose again on purpose only values of $L$
before the plateau
of the average multiplicity.
The figure gives a nice data collapse, as we expected.

\begin{figure}[htb]
\begin{center}
\resizebox{\figurewidth}{!}{\includegraphics[angle=0]{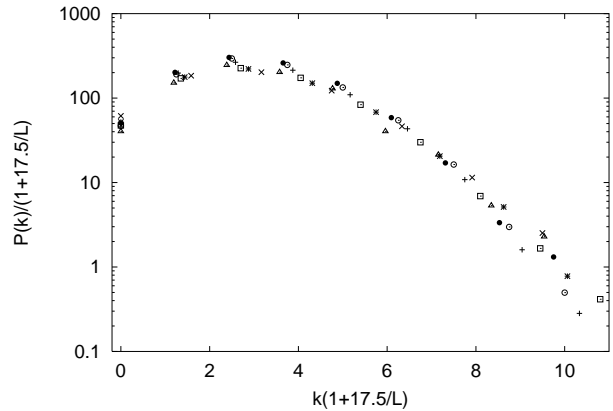}}
\end{center}
\caption{Same as Fig. \ref{histo4D}, for
five-dimensional data with MBC. One obtains again a very good
scaling of the spanning cluster multiplicity distributions for
$k>0$, by introducing the simple correction $1+17.5/L$. The lattices are:
$30^5$, $40^5$, $50^5$, $60^5$, $70^5$, $80^5$, $91^5$. The number of samples
goes from $1000$ to $50000$.}
\label{histoMBC}
\end{figure}

We repeated the analysis for the five-dimensional data, starting
from the puzzling curve of Fig. \ref{fig3}. Here we find that the
fits are very good and stable starting from the very beginning of
the curve: the fit parameters $a$ and $b$ are basically fixed
already in the range $[2:20]$; the best fit (dashed line in the
figure) was obtained including all datapoints with significant
statistics, i.e. for $2\leq L \leq 91$; we obtain $a=3.12(6)$,
$b=17.5(7)$. As one can see in Fig. \ref{fig3}, our
simple {\it ansatz} (dashed line) describes very well the observed behavior of
the data.
We notice that on the logarithmic scale for $L$
our simple scaling curve has an inflection point, exactly as the
data. This inflection can be interpreted as a signal of a possible
crossover from an initial logarithmic increase of the multiplicity
to a successive convergence to a plateau; we now see that it is
instead a natural feature of our scaling {\it ansatz}. Notice that
the correction term is more important than in four dimensions
($17.5$ vs $4.2$). That means that the data converge much more
slowly to the plateau, and explains why we could not see a
saturation even at $L=201$ (although the argument can be
reversed). Indeed, for a given $L$, the ratio $r$ of the
multiplicity to the asymptotic plateau is $r=L/(L+b)$. In 4D,
$r=96\%$ for $L=100$ and $r=98\%$ for $L=200$; in 5D one would
obtain $85\%$ and $92\%$, respectively. In order to "see" the
plateau as we do in four dimensions, we would need to go to
$L\,{\sim}\,800$! To check the consistency of the picture in 5D we
studied the scaling of the histograms $P(k,L)$, with the
correction constant $b=17.5$ that we determined above. The result
is illustrated in Fig. \ref{histoMBC}; the scaling is quite good,
except eventually for $k=0$ (but scaling laws seldom hold for
small integers) and at the very end of the tail, where the
statistics is too low and there are relevant fluctuations of the
data points.

\begin{figure}[htb]
\begin{center}
\resizebox{\figurewidth}{!}{\includegraphics[angle=0]{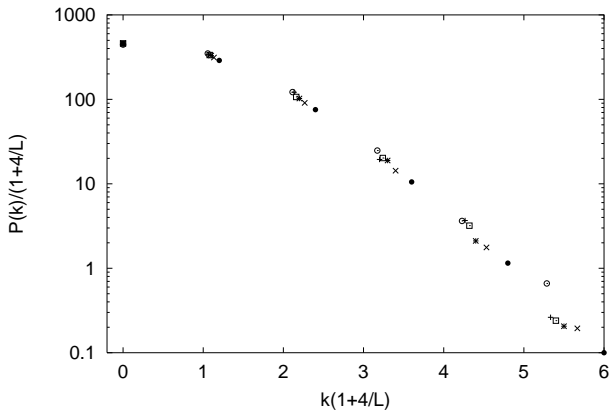}}
\end{center}
\caption{Same as Fig. \ref{histoMBC}, but with
FBC. The correction is $1+4/L$, the lattices are:
$20^5$, $30^5$, $40^5$, $50^5$, $60^5$, $70^5$. The number of samples goes from
$3000$ to $50000$.}
\label{histoFBC}
\end{figure}

\begin{figure}[htb]
\begin{center}
\resizebox{\figurewidth}{!}{\includegraphics[angle=0]{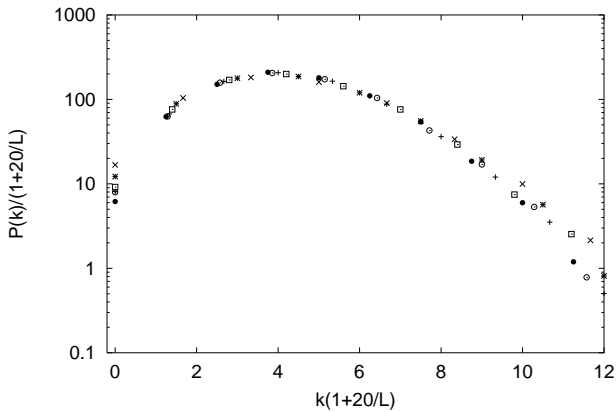}}
\end{center}
\caption{Same as Fig. \ref{histoMBC}, but with
PBC. The scaling is remarkable, probably due to the better
statistics; the necessary correction is $1+20/L$. The lattices are:
$30^5$, $40^5$, $50^5$, $60^5$, $70^5$, $80^5$. The number of samples goes from
$8000$ to $50000$.}
\label{histoPBC}
\end{figure}

Finally, we analyzed the two other data sets in five dimensions, i.e. the
ones relative to FBC and PBC. In both cases we found that our picture works:
we could find a scaling correction $X(L)=1+{\rm const}/L$ such that both
the average multiplicity and the histograms show a clean scaling. The
histograms are shown in Figs. \ref{histoFBC}
and \ref{histoPBC};
the scaling is again very good for $k>0$, with some fluctuations at the
end of the tail which are likely due to the low statistics of those points.

We also checked for five dimensions whether other percolation quantities
behave unusually, and found that they do not. The size of the largest cluster 
at the percolation threshold varies asymptotically as $L^{d-\beta/\nu}$ and the
``mean'' cluster size as $L^{\gamma/\nu}$, where $\beta/\nu \simeq 1.46, \;
\gamma/\nu \simeq 2.07$ are expected \cite{book} in five dimensions.
We found that corrections to scaling play an important
role in this range of $L$ (from 10 to 80); all our fitting curves include a
correction
term, for which we fixed the value of the exponent $\omega$ to the
estimate 0.53 given in \cite{adler2}. Taking into account this correction,
the finite size scaling fits are remarkable for all percolation
variables if we use the PBC data, for which we get
$\beta/\nu = 1.45(2)$ and $\gamma/\nu=2.08(2)$.
For MBC the fits are also very good, but a bit worse as far as the
$\chi^2$ and the values of the exponents are concerned
($\beta/\nu = 1.45(2)$, $\gamma/\nu=2.10(2)$); for the FBC
the fits are not so good and the values of the exponents not in agreement
with expectations, though quite close.
Also series expansions \cite {adler2}, which are independent of
any lattice size, gave in five dimensions the usual exponents in agreement
with expectations, without indications of particular difficulties.

\section{Conclusions}

In summary, we have derived the scaling behaviour at threshold of
the average number ${\langle}k{\rangle}$ of spanning clusters with
the linear dimension $L$ of the lattice, for any space dimension
$d$. Below the upper critical dimension $d_c=6$,
${\langle}k{\rangle}$ should approach a constant when
$L\rightarrow\infty$, for $d=6$ it should increase as $\log\,L$,
and for $d>6$ it could increase as $L^{d-6}$,
but could also
approach a constant (depending on boundary conditions and yet
unknown theoretical details).
While the latter conclusions might
seem just a confirmation of previous results on the topic, our
work highlights two new important issues:

\begin{itemize}
\item{the possibility of other scaling behaviors of
${\langle}k{\rangle}$ above the upper critical dimension, which
possibly depend on the boundary conditions;} \item{the relevance
of corrections to scaling, which may affect the scaling behaviors
up to two-three orders of magnitude in $L$.}
\end{itemize}

Our numerical investigations confirm that the multiplicity indeed
converges to a constant for $d<6$. For the case $d=5$, where we do not
see a plateau even for the largest $L$ we have taken, a simple linear correction
to scaling is able to reproduce the observed data pattern.
In six dimensions the results can be made consistent with theory
by adding a logarithmic finite-size correction; in more than six dimensions
both the data on the multiplicity $k$ and those on the order parameter
$P_{span}$ and the mean cluster size $S$ lead to very different values of the
finite-size scaling exponents
for different sets of boundary conditions, and we are not able to
derive reliable conclusions.
So, if our numerical evidence below $d_c$ is conclusive,
to close the issue above $d_
c$ simulations at much larger $L$ seem to be necessary.

\vskip1cm

\centerline{\bf ACKNOWLEDGEMENTS }

\vskip0.5cm

We acknowledge support from the German-Israeli Foundation and from
the Humboldt Foundation, P. Grassberger and J. Adler for helpful discussion
and the computer centers of the Technion (Israel;
M. Goldberg) and Julich (Germany) for time on their large
computers. S.F. acknowledges the support of the DFG Forschergruppe
under grant FOR 339/2-1. 
AC acknowledges support from
MIUR-PRIN and FIRB 2002, CRdC-AMRA,
EU Network MRTN-CT-2003-504712.


\begin{thebibliography}{99}
\bibitem{nota} We stress that this conclusion is a consequence of
considering the sum in (\ref{normP}) as an integral, which is at
best only approximately correct.
\end{thebibliography}
\end{document}